\RequirePackage{fix-cm}
\documentclass[smallextended]{svjour3}       
\smartqed  

\usepackage[english]{babel}
\usepackage{amsmath}
\usepackage{amssymb}

\input{qcircuit} 
\def\<{\langle}\def\>{\rangle}
\def\n#1{|\!|#1|\!|}\def\d{{\rm d}}

\def\Uset{\mathbb{U}}	
\def\Supp{\mathsf{Supp}\,}\def\Rnk{\mathsf{Rnk}\,}
\def\tI{\transf{I}}
\def\tA{\transf{A}}
\def\tU{\transf{U}}\def\tV{\transf{V}}\def\tT{\transf{T}}

\def\rX{\sys{X}}\def\rA{\sys{A}}\def\rB{\sys{B}}\def\rC{\sys{C}}
\def\rD{\sys{D}}\def\rI{\sys{I}}
\def\rE{\sys{E}}\def\rF{\sys{F}}

\def\trnsfrm#1{\mathcal #1}
\def\tA{\trnsfrm A} 
 
\def\tI{\trnsfrm I}\def\tT{\trnsfrm T}\def\tU{\trnsfrm U}\def\tV{\trnsfrm V} 
 \def\tZ{\trnsfrm Z}

 \def\tT{\trnsfrm T} 
\def\rA{{\rm A}}\def\rB{{\rm B}}\def\rC{{\rm C}}\def\rD{{\rm D}} \def\rE{{\rm E}} \def\rF{{\rm F}}
\def\rI{{\rm I}}

\def\rX{{\rm X}} 

\def\PurSt{\rm{PurSt}}\def\St{\rm{St}}\def\Eff{\mathrm{Eff}}
\def\Trn{\mathrm{Transf}}
\def\Rev{\mathrm{RevTrn}}

\def\Tr{\operatorname{Tr}}
\def\Hyp{{\rm Hyp}}\def\TRUE{\mathsf {TRUE}}\def\FALSE{\mathsf {FALSE}}
\def\Set{{\mathsf S}}
\def\Conv{\mathsf{Conv}}
\def\Cone{\mathsf{Cone}}

\def\trnsfrm#1{\mathcal #1}
\def\tA{\trnsfrm A} 
 
\def\tI{\trnsfrm I}\def\tT{\trnsfrm T}\def\tU{\trnsfrm U}\def\tV{\trnsfrm V} 
 \def\tZ{\trnsfrm Z}

 \def\tT{\trnsfrm T} 
\def\rA{{\rm A}}\def\rB{{\rm B}}\def\rC{{\rm C}}\def\rD{{\rm D}} \def\rE{{\rm E}} \def\rF{{\rm F}}
\def\rI{{\rm I}}

\def\rX{{\rm X}} 
\def\sH{{\mathcal{H}}}\def\sK{{\mathcal{K}}}
\def\St{\rm{St}}\def\Eff{\mathrm{Eff}}\def\Trn{\mathrm{Trn}}\def\T{\mathrm{T}}\def\Bnd{\mathrm{Bnd}}
\def\Cmplx{\mathbb{C}}\def\Reals{\mathbb{R}}
\def\CP{\mathrm{CP}}\def\P{\mathrm{P}}

\journalname{FOUNDATIONS OF PHYSICS}
\begin{document}

\title{No purification ontology, no quantum paradoxes\thanks{This work was made possible through the support of the Elvia and Federico Faggin Foundation, Grant 2020-214365.}}

\titlerunning{No purification ontology, no quantum paradoxes}        

\author{Giacomo Mauro D'Ariano}
\institute{Dipartimento di Fisica\\ dell'Universit\`a di Pavia \at
              via Bassi 6, 27100 Pavia\\
              Tel.: +39 347 0329998\\
              \email{dariano@unipv.it}\\
             \emph{Also:}  Istituto Lombardo Accademia di Scienze e Lettere\\
		INFN, Gruppo IV, Sezione di Pavia
}

\date{Received: date / Accepted: date}

\maketitle

\begin{abstract}
It is almost universally believed that in quantum theory the two following statements hold:  1) all transformations are achieved by a unitary interaction followed by a von-Neumann measurement; 2) all mixed states are marginals of pure entangled states. I name this doctrine {\em the dogma of purification ontology}.  The source of the dogma is the original von Neumann axiomatisation  of the theory, which largely relies on the Schr\H{o}dinger equation as a postulate, which holds in a nonrelativistic context, and whose operator version holds only in free quantum field theory, but no longer in the interacting theory.
\par In the present paper I prove that both ontologies of {\em unitarity} and {\em state-purity} are unfalsifiable, even in principle, and therefore axiomatically spurious. I propose instead a minimal four-postulate axiomatisation: 1) associate a Hilbert space $\sH_\rA$ to each {\em system} $\rA$; 2) {\em compose} two systems by the tensor product rule $\sH_{\rA\rB}=\sH_\rA\otimes\sH_\rB$; 3) associate a transformation from system $\rA$ to $\rB$ to a {\em quantum operation},  i.e.~ to a completely positive trace-non-increasing map between the trace-class operators of  $\rA$ and $\rB$; 4) (Born rule) evaluate all joint probabilities through that of a special type of quantum operation: the state preparation.

I then conclude that quantum paradoxes--such as the Schroedinger-cat's, and, most relevantly, the {\em information paradox}--are originated only by the dogma of purification ontology, and they are no longer paradoxes of the theory in the minimal formulation. For the same reason, most interpretations  of the theory (e.g. many-world, relational, Darwinism, transactional, von Neumann-Wigner,  time-symmetric, ...) interpret the same dogma, not the strict theory stripped of the spurious postulates.
\end{abstract}

\section{Introduction}
We all have become accustomed to a set of rules that we call {\em Quantum Theory} (QT), which we believe must hold for the whole physical domain at the fundamental level, hence also in a theory of gravity. I emphasise the naming Quantum "Theory"--instead of  Quantum "Mechanics"--to strip the rules from their mechanical instantiation in particle and matter physics. Such rules (concerning systems, states, observables, evolutions, and measurements) are now common ground for all physicists. The advent of Quantum Information has further stressed such backbone structure of the theory, with conceptual focus on composition rules of systems and transformations, hence on the underlying graph structure of QT. 

In quantum gravity the Hawking radiation from black-hole posed the problem of violation of  unitarity\cite{PhysRevD.14.2460}. This started a debate that is still open. Is it violation of unitarity an infringement of a law of QT? As a matter of facts it is almost universally believed that in QT the two following statements hold:  1) all transformations are achieved as a unitary interaction followed by a von-Neumann measurement; 2) all mixed states are marginals of pure entangled states. Such dogma of {\em ontology of purification} originated from the von Neumann axiomatisation  of QT\cite{von32}, which largely relies on the Schr\H{o}dinger equation as a postulate, the latter being valid in a nonrelativistic context and in free quantum field theory, but no longer in the interacting theory.\footnote{Indeed, besides 
being mathematically not defined, the Feynman path-integral (which in interacting field theory plays the role of the unitary evolution) in nonabelian gauge theories needs the introduction of  Faddeev and Popov {\em ghost} field modes\cite{Faddeev1967}, that, as their name says, are not experimentable.} The dogma of unitarity and purity is so widespread that we associate the nomenclature  "quantum theory of {\em open} systems" to non-unitary processes and mixed states, with a naming that emphasises the alleged incompleteness of the theoretical description.

\par In the present paper I prove that both {\em unitarity} and {\em state-purity} ontologies are not falsifiable~\cite{popp-logi,popper2002conjectures}, and therefore propose an alternative four-postulate axiomatisation of QT
: 1) associate a Hilbert space $\sH_\rA$ to each {\em system} $\rA$; 2) {\em compose} two systems by the tensor product rule $\sH_{\rA\rB}=\sH_\rA\otimes\sH_\rB$; 3) associate a transformation from system $\rA$ to $\rB$ to a {\em quantum operation},  i.e.~ to a completely positive trace-non-increasing map from $\T(\sH_\rA)$ to $\T(\sH_\rB)$; 4) provide the Born rule in terms of the probability of state-preparation--a special kind of transformation. I therefore conclude that the information paradox is not a paradox. In addition, also quantum paradoxes, such as the Schroedinger-cat's, are not logical paradoxes of the theory anymore, but just consequence of the old redundant axiomatisation. For the same reason, interpretations  of the theory as the many-world, relational, Darwinism, transactional, von Neumann-Wigner,  time-symmetric, and similia are interpretations of the dogma,  not genuine interpretations of the theory strictly speacking.

I will conclude the paper with a short discussion about the role of unitarity and state purity in the theory.

\section{The minimal and the von Neumann axiomatisations of  QT}
We assume the reader to be familiar with the natural circuit language in quantum information~\cite{Nielsen:1997p655}.\footnote{The circuit language in quantum information science is mathematically formalised in terms of the operational probabilistic theory (OPT) framework (see e.g. Ref.~\cite{CUPDCP}). Indeed, the same framework is used in computer science in terms of Category Theory\cite{CoeckePV,TullPV}.} In the following we will use the convenient rule of taking the trace $\Tr\rho$ of the density matrix $\rho\in\St(\rA)$ of system $\rA$ as the preparation probability $p(\rho)=\Tr\rho$ of the state $\rho$--our Born rule--whereas unit-trace density matrices specifically describe deterministic states. In such a way, for example,  the trace $\Tr[\tT\rho]$ is equal to the joint probability of  $\rho$-preparation followed by the quantum operation $\tT$--the composition $\tT\rho$ being just a new preparation. This convention makes possible to regard states and effects just as special cases of probabilistic transformations, from and to the trivial system $\rI$, respectively, with Hilbert space $\sH_\rI=\Cmplx$.  Finally, we will  make use of the common notation summarised in Table \ref{tnotat} in the Appendix.  

\subsection{Comparing the two axiomatisations and their main theorems}
In  Table \ref{tabminimal} we report the customary mathematical axiomatisation of QT, and the minimal axiomatisation proposed here. In both cases a system $\rA$ of the theory is mathematically associated to Hilbert space $\sH_\rA$ and the composition of systems is provided by the Hilbert-space tensor product $\sH_{\rA\rB}=\sH_\rA\otimes\sH_\rB$. It follows that the trivial system, defined by the composition rule $\rA\rI=\rI\rA=\rA$ has Hilbert space $\sH_\rI=\Cmplx$, which is the first theorem of both axiomatisations. The usage of the trivial system is crucial for considering both states and effects as special cases of transformations.

The following postulates differ remarkably between the two axiomatisations. The minimal axiomatization adds two postulates: the first one describing transformations $\tT\in\Trn(\rA\to\rB)$ by completely positive trace-not-increasing maps between trace-class operator spaces; the second one providing the Born rule in terms of the trace of a special kind of transformations corresponding to states. The axiomatisation {\em a la} von Neumann, instead, adds four independent postulates, describing: a) pure deterministic states in terms of vectors on the Hilbert space of the system; b) reversible transformations as unitary maps on states; c) a single irreversible transformation--the von Neumann-L\"{u}ders projection (on eigen-space of the "observable" corresponding to the measured "value"); d) the Born rule, providing the probability of the measured value. 
\section{The issue of unitarity: the black-hole information paradox}

The problem with the von Neumann axiomatisation, is not just a simple matter of efficiency, but what is significant is the fact that it implies that all transformations are "actually" achieved through a unitary interaction with additional systems that are not under our control, or on which we perform a von Neumann-L\"{u}ders measurement. Whereas this can be done for all transformations (and indeed it is a theorem of the minimal axiomatisation, as in Table \ref{tabminimal}), not necessarily it is actually the case. What we put into discussion here, is {\em the ontology of the unitary realisation of  quantum transformations}. 

The fact that each transformation must necessarily be ultimately unitary would be of no concern if it  made no harm to the whole logical consistency of theories in physics. However, this is not the case, due to the {\em information paradox}.  Lloyd and Preskill~\cite{Lloyd2014} expressed the impossibility of reconciling unitarity with the following relevant facts (quoting from the same reference ~\cite{Lloyd2014})
\begin{quotation}
{\em \noindent (1) An evaporating black hole scrambles quantum information without destroying it. (2) A freely falling observer encounters nothing unusual upon crossing the event horizon of a black hole. 
(3) An observer who stays outside a black hole detects no violations of relativistic effective quantum field theory.}
\end{quotation}
Then, Lloyd and Preskill say:
\begin{quotation}
{\em This puzzle has spawned many audacious ideas, beginning with Hawking's bold proposal that unitarity fails in quantum gravity. Unitarity can be temporarily violated during the black hole evaporation process, accommodating violations of monogamy of entanglement and the no-cloning principle, and allowing assumptions (1), (2), and (3) to be reconciled.}
\end{quotation}

\vfill\newpage
\begin{table}[h]
\begin{tabular}{|r| c|l|}
\hline
\multicolumn{3}{c}{\textbf{Customary mathematical axiomatisation of Quantum Theory}} \\
\hline
system &$\rA$& $\sH_\rA$\\
\hline
system composition &$\rA\rB$& $\sH_{\rA\rB}=\sH_\rA\otimes\sH_\rB$\\
\hline
deterministic pure state & $\sigma\in\PurSt_1(\rA)$ & $\sigma=|\psi\>\<\psi|$,\;$\psi\in\sH_\rA$,\;$\n{\psi}=1$\\
\hline
reversible transformations & $\tU\in\Rev(\rA)$ & $\tU\sigma=U\sigma U^\dag$, \;$U\in\Uset (\rA)$\\
\hline
\shortstack{von Neumann-L\"{u}ders\\ transformation}
 &$\sigma\to \tZ_i\sigma:=Z_i \sigma Z_i$&$\{Z_i\}_{i\in\rX}\subset\Bnd(\sH_\rA)$ PVM\\ 
\cline{1-2}
Born rule &$p(i|\psi)=\<\psi|Z_i|\psi\>$&
\\ 
\hline
\multicolumn{3}{c}{\textbf{Theorems}} \\
\hline
trivial system &$\rI$& $\sH_\rI=\Cmplx$\\
\hline
deterministic states& $\rho\in\St_1(\rA)\equiv\Conv(\PurSt_1(\rA))$ & $\rho\in\T^+_{=1} (\sH_\rA)$\\
\hline
states& $\rho\in\St(\rA)\equiv\Cone_{\leqslant 1}(\PurSt_1(\rA))$ & $\rho\in\T^+_{\leqslant1} (\sH_\rA)$\\
\hline
\shortstack{\\Transformation as\\ unitary interaction\\+ von Neumann\\ observable on ``meter''}
  & 
  \shortstack{   $\!\!\!\begin{aligned}
 \Qcircuit @C=1em @R=.7em @! R {
        &\poloFantasmaCn{\rA}\qw&\gate{\tT_i}&\poloFantasmaCn{\rB}\qw&\qw}
    \end{aligned} =\!\!\!
    \begin{aligned}
  \Qcircuit
    @C=1em @R=.7em @! R {
&\poloFantasmaCn{\rA}\qw&\multigate{1}{\tU}  &\poloFantasmaCn{\rB}\qw&\qw\\
\prepareC{\sigma}  &\poloFantasmaCn{\rF}\qw&\pureghost{\tU}\qw&\poloFantasmaCn{\rE}\qw&\measureD{Z_i}}
\end{aligned}$}
  &
  \shortstack{$\tT_i\rho=\Tr_\rE[U(\rho\otimes\sigma)U^\dag(I_\rB\otimes Z_i)]$} 
\\
\hline
transformation & $\tT\in\Trn(\rA\to\rB)$ & $\tT\in\CP_{\leqslant}(\T(\sH_\rA)\to\T(\sH_\rB))$\\
\hline
parallel composition &  $\tT_1\in\Trn(\rA\to\rB)$, $\tT_2\in\Trn(\rC\to\rD)$ &$\tT_1\otimes\tT_2$ \\
\hline
sequential composition &  $\tT_1\in\Trn(\rA\to\rB)$, $\tT_2\in\Trn(\rB\to\rC)$ &$\tT_2\tT_1$ \\
\hline
effects& $\epsilon\in\Eff(\rA)\equiv\Trn(\rA\to\rI)$ & $\epsilon(\cdot)=\Tr_\rA[\cdot E],\; 0\leqslant E\leqslant I_A$\\
\hline
  & $\epsilon\in\Eff_1(\rA)\equiv\Trn_1(\rA\to\rI)$ & $\epsilon=\Tr_\rA$\\
\hline
\multicolumn{3}{c}{\textbf{Minimal mathematical axiomatisation of Quantum Theory}} \\
\hline
system &$\rA$& $\sH_\rA$\\
\hline
system composition &$\rA\rB$& $\sH_{\rA\rB}=\sH_\rA\otimes\sH_\rB$\\
\hline
transformation & $\tT\in\Trn(\rA\to\rB)$ & $\tT\in\CP_{\leqslant}(\T(\sH_\rA)\to\T(\sH_\rB))$\\
\hline
Born rule &$p(\tT)=\Tr\tT$& $\tT\in\Trn(\rI\to\rA)$\\
\hline
\multicolumn{3}{c}{\textbf{Theorems}} \\
\hline
trivial system &$\rI$& $\sH_\rI=\Cmplx$\\
\hline
reversible transformation & $\tU=U\cdot U^\dag$ & $U\in\Uset(\sH_\rA)$\\
\hline
determ. transformation & $\tT\in\Trn_1(\rA\to\rB)$ & $\tT\in\CP_{=}(\T(\sH_\rA)\to\T(\sH_\rB))$\\
\hline
parallel composition &  $\tT_1\in\Trn(\rA\to\rB)$, $\tT_2\in\Trn(\rC\to\rD)$ &$\tT_1\otimes\tT_2$ \\
\hline
sequential composition &  $\tT_1\in\Trn(\rA\to\rB)$, $\tT_2\in\Trn(\rB\to\rC)$ &$\tT_2\tT_1$ \\
\hline
states& $\rho\in\St(\rA)\equiv\Trn(\rI\to\rA)$ & $\rho\in\T^+_{\leqslant1} (\sH_\rA)$\\
\hline
  & $\rho\in\St_1(\rA)\equiv\Trn_1(\rI\to\rA)$ & $\rho\in\T^+_{=1} (\sH_\rA)$\\
\hline
          & $\rho\in\St(\rI)\equiv\Trn(\rI\to\rI)$ & $\rho\in[0,1]$\\
\hline
          & $\rho\in\St_1(\rI)\equiv\Trn(\rI\to\rI)$ & $\rho=1$\\
\hline
effects& $\epsilon\in\Eff(\rA)\equiv\Trn(\rA\to\rI)$ & $\epsilon(\cdot)=\Tr_\rA[\cdot E],\; 0\leqslant E\leqslant I_A$\\
\hline
  & $\epsilon\in\Eff_1(\rA)\equiv\Trn_1(\rA\to\rI)$ & $\epsilon=\Tr_\rA$\\
\hline
\shortstack{\\Transformations as \\ unitary interaction\\ +\\ von Neumann-L\"{u}ders}
    &
 $ \!\!\!\begin{aligned}
      \Qcircuit @C=1em @R=.7em @! R {
        &\poloFantasmaCn{\rA}\qw&\gate{\tT_i}&\poloFantasmaCn{\rB}\qw&\qw}
    \end{aligned} =\!\!\!
    \begin{aligned}
  \Qcircuit
    @C=1em @R=.7em @! R {
&\poloFantasmaCn{\rA}\qw&\multigate{1}{\tU}  &\poloFantasmaCn{\rB}\qw&\qw\\
\prepareC{\sigma}  &\poloFantasmaCn{\rF}\qw&\pureghost{\tU}\qw&\poloFantasmaCn{\rE}\qw&\measureD{Z_i}
}
\end{aligned}$\!\!
& 
  \shortstack{$\tT_i\rho=\Tr_\rE[U(\rho\otimes\sigma)U^\dag(I_\rB\otimes Z_i)]$}
\\
\hline
\end{tabular}
\caption{Customary versus minimal mathematical axiomatisation of Quantum Theory}\label{tabminimal}
\end{table}
\newpage

On the other hand, Nikolic writes that "violation of unitarity by Hawking radiation does not violate energy-momentum conservation"~\cite{Nikoli_2015}, hence it makes no harm to physics. 

Antonini and Nambiar write\cite{Antonini2018TheBH}
\begin{quotation}
{\em This is the essence of the black hole information paradox (BHIP): unlike any other classical or quantum system, black holes may not conserve information, thus violating unitarity.
Some physicists speculate that quantum gravity may actually be non-unitary.

When this phenomenon is analyzed closer, we discover that it takes pure states to mixed states, a violation of unitarity, a fundamental property of quantum physics.}
\end{quotation}

And Polchinski declared~\cite{polchinskytalk}: 
\begin{quotation}
{\em Unitarity? Not consistent with AdS/CFT.}
\end{quotation}

In the following sections we will see that {\em unitarity of the realisation of quantum transformations is a spurious postulate}, since in addition to be inessential, it is also not falsifiable. The same holds for the requirement of state purity as the actual realisation of mixed states as marginal of pure entangled  ones, as in most interpretations of quantum theory, e.g. the many-world. With these motivations we devote the entire next section to develop the theory of quantum falsification, and apply it to prove unfalsifiability of purity of quantum states, unitarity of quantum transformations, and consequently the unfalsifiability of unitary realisation of transformations and pure realisation of mixed states. 

\section{The quantum falsification test}
\begin{definition}[Falsifier] The event $F$ is a {\em falsifier} of hypothesis $\Hyp$ if $F$ cannot happen for $\Hyp=\TRUE$. 
\end{definition}
Accordingly we will call the binary test $\{F,F_?\}$ {\em a falsification test} for hypothesis $\Hyp$, $F_?$ denoting the {\em inconclusive event}.\footnote{We want to remark that the occurrence of $F_?$ generally does not mean that $\Hyp=\TRUE$, but only that the falsification test failed.} Practically one is interested in {\em effective} falsification tests $\{F,F_?\}$ which are not singleton--the two singleton tests corresponding to the {\em inconclusive falsification test} for $F=0$  and the {\em logical falsification} for $F_?=0$, respectively.

\medskip
Suppose now that one wants to falsify a proposition about the state $\rho\in\St(\rA)$ of system $\rA$. In such case any effective falsification test  can be achieved as a binary {\em observation test} of the form
\begin{equation}
\{F,F_?\}\subset\Eff(A),\quad F_?:=I_\rA-F,\quad F>0,F_?\geq 0,
\end{equation}
where with the symbol $F$ ($F_?$) we denote both the event  and its corresponding positive operator. Notice the strict positivity of $F$ for effectiveness of the test,  $F=0$ corresponding to the {\em inconclusive test}, namely the test that outputs only the inconclusive outcome. On the other hand, the case $F_?=0$ corresponds to logical {\em a priori} falsification. 
\subsubsection*{Example of falsification test}
Consider the proposition
\begin{equation}\quad\label{assert}
\Hyp:\quad\Supp\rho=\sK\subset\sH_\rA,\;\rho\in\St(\rA),\qquad\dim\sH_\rA\geq 2
\end{equation}
where $\Supp\rho$ denotes the support of  $\rho$. Then,  any operator of the form
\begin{equation}\label{falsifier}
0<F\leqslant I_\rA,\quad \Supp F\subseteq\sK^\perp
\end{equation}
would have zero expectation for a state $\rho$ satisfying $\Hyp $ in Eq. \eqref{assert}, which means that occurrence  of $F$ would be a falsification of  \Hyp, namely
\begin{equation}
\Tr[\rho F]>0\,\Rightarrow \Hyp=\FALSE.
\end{equation}
In this example we can see how the falsification test is not dichotomic, namely the occurrence of $F_?$ does not mean that $\Hyp=\TRUE$, since $F_?$ occurs if $\Supp F_?\cap\sK\neq 0$. Eq. \eqref{falsifier} provides the most general falsification test of  $\Hyp$ in Eq. \eqref{assert}, and the choice $\Supp F=\sK^\perp$ provides the most efficient test since it maximises the falsification chance.

\medskip
We may have considered more generally falsification tests with $N\geq1$ falsifiers and $M\geq 1$ inconclusive events. However, any of such a test would correspond to a set of  binary falsification tests with the falsifier made as coarse-graining of falsifiers only, and among such tests the most efficient one being the one which coarse-grains all falsifiers into a single falsifier and all inconclusive events into a single inconclusive event. Another relevant observation is that, by {\em modus tollens}, if $\Hyp_1\Rightarrow\Hyp_2 $ a falsifier for $\Hyp_2$ also falsifies  $\Hyp_1$.

\smallskip
In the following section we will see that unitarity of the realisation of quantum transformations is actually a spurious postulate, since in addition to be inessential, it is also not falsifiable. We devote the entire next section to quantum falsification theory and apply it to prove unfalsifiability of purity of quantum states, and unitarity of quantum transformations, and consequently the unfalsifiability of unitary realisation of transformations and pure realisation of mixed states. 

\section{Unfalsifiabilities in quantum theory}
We will now  prove a set of no-falsification theorems within quantum theory.  

\subsection{Unfalsifiability of purity of a quantum state}
\begin{theorem}[Unfalsifiability of state purity]\label{falspurST} There exists no test falsifying purity of an unknown state of a given system  $\rA$.
\end{theorem}
\proof
In order to falsify the hypothesis 
\begin{equation}\label{assertpur}
\Hyp:\; \rho\in\PurSt(\rA),
\end{equation}
we need a falsifier $F\in\Eff(\rA)$ satisfying
\begin{equation}
\Tr[F\rho]=0,\;\forall\rho\in\PurSt(\rA),
\end{equation}
which means that 
\begin{equation}
\forall\psi\in\sH_\rA:\,\<\psi|F|\psi\>=0,
\end{equation}
namely $F=0$, which means that the test is inconclusive.\qed

By the same argument one can easily prove the impossibility of falsifying purity even when $N>1$ copies of the state are available.

\subsection{Unfalsifiability of atomicity of a quantum transformation}
The impossibility of falsifying purity of a state has as  an immediate consequence the impossibility of falsifying the atomicity of a transformation.\footnote{A transformation is atomic, namely non refinable non trivially, when it has only one Krauss operator in its Krauss form. Equivalently, its Choi-Jamiolkowsky operator is rank-one.}

\begin{theorem}[Unfalsifiability of transformation atomicity] There exists no test falsifying atomicity of an unknown transformation
$\tA\in\Trn(\rA\to\rB)$.
\end{theorem}
\proof The most general scheme for testing a property of a transformation $\tT\in\Trn(\rA\to\rB)$ is the following
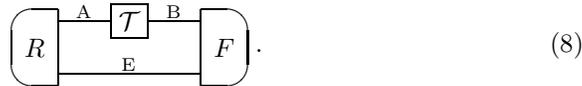
\begin{equation}\label{scheme}
\begin{aligned}
    \Qcircuit @C=1em @R=.7em @! R {\multiprepareC{1}{R}&\poloFantasmaCn{\rA}\qw&\gate{\tT}&\poloFantasmaCn{\rB}\qw&\multimeasureD{1}{F}\\
      \pureghost{R}&\qw&\poloFantasmaCn{\rE}\qw&\qw&\ghost{F}}
\end{aligned}\;.
  \end{equation}
We can use the maximally entangled state $R=|\Phi\>\<\Phi|$, thus exploiting the Choi-Jamio\l{}kowski cone-isomorphism between transformations and bipartite states. One has
\begin{equation}
\text{\em atomicity of }\tT\equiv\text{\em purity of  state }(\tT\otimes\tI_\rE) R,
\end{equation}
and falsifying atomicity of $\tT\in\Trn(\rA\to\rB)$ is equivalent to falsifying purity of 
$(\tT\otimes\tI_\rE) R$, which is impossible.\qed

\subsection{Unfalsifiability of max-entanglement of a pure bipartite state}
In the following we will use maximally entangled pure bipartite states in $\sH_\rA\otimes\sH_\rB$ generally with non equal dimensions $d_\rA\geq d_\rB$ and Schmidt number equal to $d_\rB$. A maximally entangled state of this kind has the general form 
\begin{equation}\label{doppioket}
|V\>=\sum_{n=1}^{d_\rA}\sum_{m=1}^{d_\rB}V_{nm}|n\>\otimes|m\>,
\end{equation}
where the matrix of coefficients $V_{nm}$ correspond to the isometry 
\begin{equation}
V=\sum_{n=1}^{d_\rA}\sum_{m=1}^{d_\rB}V_{nm}|n\>\<m|,
\end{equation}
satisfying $V^\dag V=I_\rB$.
We are now in position to prove the following theorem.
\begin{theorem}[Unfalsifiability of max-entanglement of a pure state of systems $\rA\rB$]\label{falsU} There exists no test falsifying max-entanglement of a pure bipartite state.
\end{theorem}
\proof W.l.g. we consider the case of $d_\rA\geq d_\rB$, as in Eq. \eqref{doppioket}. Falsification of max-entanglement of state $|V\>\<V|$ needs a falsifier $F\in\Eff(\rA\rB)$ satisfying
\begin{equation}
\Tr[F|V\>\<V|]=0,\;\forall |V\>\text{ maximally entangled}.
\end{equation}
In particular, since unitary transformations on either $\sH_\rA$ or $\sH_\rB$ preserve max-entanglement, one has
\begin{equation}\label{max-entang}
\Tr[F(\tU\otimes\tI_\rB)|V\>\<V|]=0,\;\forall\tU=U\cdot U^\dag, U\in\Uset(\sH_\rA).
\end{equation}
It follows that the average over the unitary group $G_\rA=SU(d_\rA)$ must be zero, corresponding to\footnote{In Eq. \eqref{Gav} $\d\tU$ is denotes the invariant normalized Haar measure of $G_\rA=SU(d_\rA)$.}
\begin{equation}\label{Gav}
\begin{split}
0=&\int_{G_\rA}d \tU\Tr[F(\tU\otimes\tI_\rB)|V\>\<V|]=\Tr[F(I_\rA\otimes \Tr_\rA[|V\>\<V|])]\\=&\Tr[F (I_\rA\otimes ( V^\dag V)^*)]=\Tr[F (I_\rA\otimes I_\rB)]=\Tr[F],
\end{split}
\end{equation}
where the complex conjugation is w.r.t. the chosen basis in Eq.\eqref{doppioket}. Eq. \eqref{Gav} implies that $F=0$, which contradicts the falsification effectiveness condition $F>0$.\qed

%
\subsection{Unfalsifiability of isometricity a quantum transformation}
\begin{theorem}[Unfalsifiability of isometricity of a transformation from $\rB$ to $\rA$ with $d_\rA\geq d_\rB$]\label{falsV} There exists no test falsifying isometricity of a transformation $\tV\in\Trn(\rB\to\rA)$ with $\dim\sH_\rA\geq\dim\sH_\rB$.
\end{theorem}
%
%
%
\proof The application of the operator to a fixed maximally-entangled state puts isometricity transformations in one-to-one correspondence with maximally entangled states. Thus, falsifying maximal isometricity on a transformation is equivalent to falsifying maximal entanglement of a state, which is impossible.\qed

\begin{corollary}[It is not possible to falsify unitarity of a transformation]
\end{corollary}
\proof Obviously Theorem \ref{falsV} exclude the possibility of falsifying unitarity of a transformation, since it is a special case of isometricity.

\smallskip
\subsection{Unfalsifiability of a mixed state being the marginalization of a pure one}
Any purification of the mixed state $\rho\in\St(\rA)$  can be written in the following diagrammatic form
\begin{equation}
\begin{aligned}\label{Pur}
    \Qcircuit @C=1em @R=.7em @! R {\prepareC{\rho}&\qw\poloFantasmaCn{\rA}\qw&\qw}
    \end{aligned}    ~=~
\begin{aligned}
    \Qcircuit @C=1em @R=.7em @! R {\multiprepareC{1}{\rho^{1/2}}&\poloFantasmaCn{\rA}\qw&\qw&\qw&\qw\\
      \pureghost{\rho^{1/2}}&\poloFantasmaCn{\rB}\qw&\gate{\tV}&\poloFantasmaCn{\rE}\qw&\measureD{e}}
  \end{aligned},
\end{equation}
with $d_\rB=d_\rA\leqslant d_\rE$ and $e$ denoting the deterministic effect, corresponding to discarding system $\rE$, and $\tV$ being any map isometric on $\Supp\rho$. We thus resort to the falsifiability of being a pure state of the form $(I_\rA\otimes V)|\rho^{1/2}\>_{\rA\rA}$. 
\begin{theorem}[Unfalsifiability of mixed state in $\St(\rA)$ being the marginalisation of a pure state of $\rA\rE$ with $d_\rE\geq d_\rA$]\label{falspuri} There exists no test falsifying the assertion that a mixed state in $\St(\rA)$ is actually the marginal of a pure state  of  $\rA\rE$ with $\dim\sH_\rE\geq\dim\sH_\rA$.
\end{theorem}
\proof
Consider the general purification scheme in Eq. \eqref{Pur}. Upon denoting by $\tV\in\Trn(\rA\to\rE)$ an isometric transformation with $d_\rE\geq d_\rB=d_\rA$, a falsifier $F\in\Bnd^+(\rA\rE)$ should satisfy the following identity
\begin{equation}
\Tr[F(\tI_\rA\otimes\tV)|\rho^{1/2}\>\<\rho^{1/2}|]=0,
\end{equation}
and by unitarily connecting all the possible isometries $\tV$ with fixed support, one has
\begin{equation}
\Tr[F(\tI_\rA\otimes\tU\tV)|\rho^{1/2}\>\<\rho^{1/2}|]=0,\quad\forall\tU,\; \tU=U\cdot U^\dag,\; U\in\Uset(\sH_\rE).
\end{equation}
It follows that the average over the unitary group $G_\rE=SU(d_\rE)$ must be zero, corresponding to
\begin{equation}\label{Gav2}
\begin{split}
0=&\int_{G_\rE}d \tU\Tr[F(\tI_\rA\otimes\tU\tV)|\rho^{1/2}\>\<\rho^{1/2}|]
=\Tr[F(\Tr_\rE[|\rho^{1/2}V^T\>\<\rho^{1/2}V^T|]\otimes I_\rE ])]\\=&\Tr[F (\rho\otimes I_\rE)]=\Tr[F_\rA\rho],
\end{split}
\end{equation}
where $T$ denotes the transpose w.r.t. the basis for the representation of the $\rho^{1/2}$ purification, and $F_\rA=\Tr_\rE F$. For $\rho$ full-rank one has $F_\rA=0$, implying $\Tr F=0$, namely $F=0$, proving the statement. For $\Rnk\rho<d_\rA$, $F_\rA$ becomes a falsifier of $\Supp\rho$, which is known a priori.\qed

This excludes the possibility of falsifying that a knowingly mixed state of a quantum system $\rA$ is actually the marginal of a pure entangled state with an environment system $\rE$. Moreover, the system $\rE$ is unknown (we just know that it must have dimension $d_\rE\geq d_\rA$).
\subsection{Unfalsifiability of unitary realization of a transformation}
The impossibility of falsifying the unitarity of a transformation (Theorem \ref{falsU}) with input and output systems under our control excludes the possibility of falsifying that a transformation is actually achieved unitarily, according to the scheme
\begin{equation}
\begin{aligned}
      \Qcircuit @C=1em @R=.7em @! R {
        &\poloFantasmaCn{\rA}\qw&\gate{\tT_i}&\poloFantasmaCn{\rB}\qw&\qw}
    \end{aligned}~=~
 \begin{aligned}
  \Qcircuit
    @C=1em @R=.7em @! R {
&\poloFantasmaCn{\rA}\qw&\multigate{1}{\tU}  &\poloFantasmaCn{\rB}\qw&\qw\\
\prepareC{\sigma}  &\poloFantasmaCn{\rF}\qw&\pureghost{\tU}\qw&\poloFantasmaCn{\rE}\qw&\measureD{Z_i}
}
\end{aligned},
\end{equation}
with $\{Z_i\}$ von Neuman-L\"{u}ders measurement over the output environment $\rE$, and the input environment $\rF$  prepared in a state $\sigma$.
Systems $\rE, \rF$, state $\sigma$, measurement $Z_i$, and unitary $\tU$ are all not unique and unknown, otherwise  the testing resorts to falsifying unitarity of $\tU$, which is impossible, not even with control of input-output systems $\rA\rF$ and $\rB\rE$.

\section{Conclusions}
Some authors argue that unobservable physics (e.~g. cosmological models invoking a multiverse) is legitimate scientific theory, based on abduction and empirical success\cite{carroll2018falsifiability}. However, I think that we should keep cosmology as an exception. Quantum Theory should be taken at a completely different level of consideration. It is a mature theory, it is under  lab control, and, by its own nature, it categorises the same rules for experiments. For such a theory, {\em falsifiability}, at least {\em in principle}, is a necessary requirement. The case of unitarity and the information paradox is paradigmatic in this respect, and one may legitimately ask what is the point in keeping within the theory an inessential metaphysical statement, without which the theory perfectly stands on its own legs.
Somebody may argue that unitarity is dictated by a more refined theory, e.~g. quantum field theory. However, as discussed in this paper, although this is the case for the free theory, it no longer survives the interacting one.

If not falsifiable and inessential, why then unitarity is so relevant to the theory? Why vectors in Hilbert spaces are ubiquitous? The answer is that unitarity and purity are powerful symmetries of the theory, and, as such, they play a crucial role in theoretical evaluations. \par
Finally, we have noticed that most interpretations of the theory (many-world, relational, Darwinism, transactional, von Neumann-Wigner,  time-symmetric, ...) are indeed interpretations of the purification-ontology dogma---not genuine interpretations of the theory strictly speaking. Such interpretations, however, still play a role as models, helping our conceptual understanding and intuition. However, they should not be taken too seriously. This is the main lesson of Copenhagen.
\appendix
\section{Appendices}\vskip10pt
\subsection{Notation}
\begin{table}[hb]
\begin{tabular}{|l|l|}
\hline
$\sH$ & Hilbert space over $\Cmplx$\\
$\Bnd^+(\sH)$  &bounded positive operators over $\sH$\\
$\Uset(\sH)$ &unitary group over $\sH$\\
$\T(\sH)$ &trace-class operators over $\sH$\\
$\T^+(\sH)$  &trace-class positive operators over $\sH$\\
$\T_{\leqslant1}^+(\sH)$ &positive sub-unit-trace operators over $\sH$\\
$\T_{=1}^+(\sH)$ &positive unit-trace operators over $\sH$\\
$\CP_{\leqslant}$ &trace-non increasing completely positive map\\
$\CP_{=}$ &trace-preserving completely positive map\\
$\Conv(\Set)$ &convex hull of $\Set$\\
$\Cone(\Set)$ &conic hull of $\Set$\\
$\Cone_{\leqslant 1}(\Set)$ &convex hull of $\{\Set\cup{0}\}$\\
$\St(\rA)$ &set of states of system $\rA$\\ 
$\St_1(\rA)$& set of deterministic states of system $\rA$\\ 
$\Eff(\rA)$& set of effects of system $\rA$ \\
$\Eff_1(\rA)$& set of deterministic effects of system $\rA$ \\
$\Trn(\rA\to\rB)$& set of transformations from system $\rA$ to system $\rB$  \\
$\Trn_1(\rA\to\rB)$& set of deterministic transformations from system $\rA$ to system $\rB$  \\
\hline
& {\bf Special cases}\\
\hline
&$\T(\Cmplx)=\Cmplx$,\;\;   $\T^+(\Cmplx)=\Reals^+$,\;\; $\T_{\leqslant1}^+(\Cmplx)=[0,1]$,\;\; $\T_{=1}^+(\Cmplx)=\{1\}$\\
&$\CP(\T(\sH)\to\T(\Cmplx))=\P(\T(\sH)\to\T(\Cmplx))=\{\Tr[\cdot E],\, E\in\Bnd^+(\sH)\}$
\\
&$\CP(\T(\Cmplx)\to\T(\sH))=\P(\T(\Cmplx)\to\T(\sH))=\T^+(\sH)$
\\
&$\CP_{\leqslant}(\T(\Cmplx)\to\T(\sH))\equiv\T^+_{\leqslant 1}(\sH)$\\
&$\CP_{\leqslant}(\T(\sH)\to\T(\Cmplx))\equiv\{\epsilon(\cdot)=\Tr[\cdot E],\, 0\leqslant E\leqslant I\}$\\
\hline
\end{tabular}
\caption{Notation and  corollary special cases.}\label{tnotat}
\end{table}
\begin{acknowledgements}
I thank Alessandro Tosini for discussions and a careful reading of the manuscript. I also thank Mio Murao for suggesting me to generalize to isometries the no-falsification theorem for unitaries. I finally express my best thanks to the anonymous Referee for pointing me out that Blake C. Stacey's paper \cite{Stacey2016} contains the following quotation from Chris Fuchs highly sympathetic with the ideas of the present paper
\begin{quote}
{\em Von Neumann's setting the issue of measurement in these terms was the great original sin of the quantum foundational debate.}
\end{quote}
\end{acknowledgements}


\end{document}